\documentstyle[11pt,aaspp4]{article}

\def\etal{{\sl et al.\/}\ }

\lefthead{Shen et al.}
\righthead{Southern Hemisphere VLBI SURVEY - II}

\begin{document}

\title{A 5-GHz Southern Hemisphere VLBI Survey of \\
       Compact Radio Sources - II}
\vspace*{-0.35cm}

\author{Z.-Q. Shen\altaffilmark{1}}
\affil{Harvard-Smithsonian CfA, 60 Garden Street, Cambridge, Massachusetts 02138 \\
       and Shanghai Astronomical Observatory, 80 Nandan Road, Shanghai 200030, China}

\author{T.-S. Wan}
\affil{Shanghai Astronomical Observatory, 80 Nandan Road, Shanghai 200030, China}
 
\author{J. M. Moran}
\affil{Harvard-Smithsonian CfA, 60 Garden Street, Cambridge, MA 02138}

\author{D.~L.~Jauncey,~J.~E.~Reynolds,~A.~K.~Tzioumis,~R.~G.~Gough,~R.~H.~Ferris~and~M.~W.~Sinclair}
\affil{Australia Telescope National Facility, CSIRO, Epping, NSW~2121, Australia}
 
\author{D.-R. Jiang, X.-Y. Hong and S.-G. Liang}
\affil{Shanghai Astronomical Observatory, 80 Nandan Road, Shanghai 200030, China}

\author{P. G. Edwards}
\affil{The Institute of Space and Astronautical Science, Sagamihara, Kanagawa 229, Japan}

\author{M. E. Costa\altaffilmark{2}}
\affil{Physics Department, University of Western Australia, Nedlands, WA 6009, Australia} 

\author{S. J. Tingay\altaffilmark{3}}
\affil{Mount Stromlo and Siding Spring Observatories, ACT 2611, Australia} 

\author{P. M. McCulloch, J. E. J. Lovell\altaffilmark{4} and E. A. King\altaffilmark{5}}
\affil{Department of Physics, University of Tasmania, Hobart, Tasmania 7001, Australia}

\author{G. D. Nicolson}
\affil{Hartebeesthoek Radio Astronomy Observatory, Krugersdorp 1740, South Africa}

\author{D. W. Murphy, D. L. Meier, and T. D. van Ommen\altaffilmark{6}}
\affil{Jet Propulsion Laboratory, California Institute of Technology, Pasadena, California 91109}

\author{G. L. White}
\affil{Physics Department, University of Western Sydney, Nepean, NSW 2747, Australia}

\altaffiltext{1}{Present address: Institute of Astronomy and Astrophysics, Academia Sinica,
                 P.O. Box 1-87, Nankang, Taipei 115, Taiwan}
\altaffiltext{2}{Present address: University of Tasmania, Hobart, Tasmania 7001, Australia}
\altaffiltext{3}{Present address: Jet Propulsion Laboratory, California Institute of Technology,
                 Pasadena, California 91109}
\altaffiltext{4}{Present address: The Institute of Space and Astronautical Science,
                 Sagamihara, Kanagawa 229, Japan}
\altaffiltext{5}{Present address: Australia Telescope National Facility, CSIRO, Epping,
                 NSW~2121, Australia}
\altaffiltext{6}{Present address: Antarctic CRC, University of Tasmania, G.P.O. 252C
                 Hobart, Tasmania, Australia}
\clearpage

\begin{abstract}
We report the results of a 5-GHz southern-hemisphere snapshot VLBI observation
of a sample of blazars. The observations were performed with the Southern
Hemisphere VLBI Network plus the Shanghai station in 1993 May. Twenty-three
flat-spectrum, radio-loud sources were imaged. These are the first VLBI images
for 15 of the sources. Eight of the sources are EGRET ($>$~100~MeV)
$\gamma$-ray sources. The milliarcsecond morphology shows a core-jet structure
for 12 sources, and a single compact core for the remaining 11. No compact
doubles were seen. Compared with other radio images at different epochs and/or 
different frequencies, 3 core-jet blazars show evidence of bent jets, 
and there is some evidence for superluminal motion in the cases of 2 blazars. 
The detailed descriptions for individual blazars are given. This is the second
part of a survey: the first part was reported by Shen \etal (1997).
\end{abstract}

\section{Introduction}
Blazar is the collective name for BL Lac objects, optically violent
variables and highly polarized quasars, all of which share extreme
observational properties that distinguish them from other active
galactic nuclei. These properties include strong and rapid
variability, high optical polarization, weak emission lines, and
compact radio structure (cf. Impey 1992). About 200 blazars
have been identified (cf. Burbidge \& Hewitt 1992). A possible
explanation for the blazar phenomenon within a unified scheme for
active galactic nuclei is that their emission is beamed by the
relativistic motion of the jets traveling in a direction close to
the observer's line of sight. 
This beaming argument is strengthened by the recent CGRO (Compton Gamma Ray
Observatory) discovery that most of the detected high-latitude  
$\gamma$-ray sources are blazars (e.g. Dondi \& Ghisellini 1995).
A comprehensive theoretical review of these sources
has been made by Urry \& Padovani (1995).

Blazars are an important class of active galactic nuclei because they
are thought to be sources with relativistic jets seen nearly
end-on. Such sources generally have very compact, flat-spectrum radio
cores, which are appropriate for VLBI study.  Pearson \& Readhead
(1988) have undertaken a survey of a complete sample consisting of 65 strong
northern-hemisphere radio sources. They provided the first well-defined
morphological classification scheme, based primarily on the large-scale radio
structure and radio spectra of the sources. 

Most surveys to date, however, including the recent Caltech--Jodrell Bank VLBI
Surveys (Polatidis \etal 1995; Thakkar \etal 1995; Xu \etal 1995; Taylor \etal
1994; Henstock \etal 1995), have been restricted to northern-hemisphere
sources. For example, all the confirmed superluminal radio sources, except the
well-known equatorial source 3C~279 (1253$-$055), are in the northern sky
(Vermeulen \& Cohen 1994). This reflects the paucity of southern VLBI
observations, the notable exceptions being the systematic one-baseline surveys
by Preston \etal (1985) and Morabito \etal (1986), and the more extensive SHEVE
survey (Preston \etal 1989 and references therein).

Since 1992 we have been carrying out a program to address this deficiency,
using VLBI at 5 GHz to study southern radio sources. In an earlier paper
(Shen \etal 1997, hereafter Paper I), we reported the results from the first
observing session in 1992 November, and presented images of 20 strong sources
selected on the basis of their correlated fluxes on intercontinental baselines. 
In 1993 May we observed a second sample of southern sources, which is the
subject of this paper.
 
Section~2 introduces this blazar sample; Section~3 briefly describes the
observations and data reduction procedures; Section~4 presents
the results; the summary and conclusions are presented in Section~5.

Throughout the paper, we define the spectral index, $\alpha$, by the
convention {S$_{\small \nu}\,\propto\,{\nu}^{\alpha}$}, and assume
{H$_0=100$~km~s$^{-1}$~Mpc$^{-1}$} and q$_0=0.5$. 

\section{The Blazar Sample}

We selected our sample of southern blazars from Table~1 of Burbidge \&
Hewitt (1992), which was based on the 2.7-GHz Parkes Survey of Bolton \etal
(1979) and the 5-GHz survey of K\"uhr \etal (1981). The
sample is defined by the following criteria:
\begin{enumerate}
\item{ Declination: $-55^\circ < \delta < -10^\circ$, }
\item{ total flux density at 5.0~GHz: S$^{\small t}_{\small 5.0GHz} >$1.0~Jy, }
\item{ radio spectral index between 2.7 and 5.0~GHz:
$\alpha^{\small 5.0~GHz}_{\small 2.7~GHz}~\geq$~$-$0.5.}
\end{enumerate}
Of the 218 sources listed by Burbidge \& Hewitt (1992), 24 meet the above
criteria. These include PKS~1334--127, PKS~1504--166 and PKS~1519--273, 
which were observed in
the first session (Paper I) and not re-observed here. 
The remaining 21 sources are listed in Table~1, along with 2 additional
sources that were included in the observing run: 
3C~273 (1226+023) and PKS~1127--145, a radio-loud quasar.
PKS~0823$-$223 has the lowest galactic latitude of this sample, with
$b \sim 9^\circ$; for all other sources, $|b| > 17^\circ$.

Seven blazars (PKS~0332--403, PKS~0426--380, PKS~0438--436, PKS~0521--365,
PKS~1034--293, PKS~1226+023 and PKS~2234--123) belong to the 36-source sample
described in Paper I, but were not observed in 1992 November.  Also, 6 sources
(PKS~0208$-$512, PKS~0521$-$365, PKS~0537$-$441, PKS~1127$-$145,
PKS~1226+023 and PKS~1424$-$418) have been detected by the EGRET
(Energetic Gamma-Ray Experiment Telescope) on board the 
CGRO (Mattox \etal 1997), and two blazars (PKS~0454$-$234 and 
PKS~2005$-$489) were marginally detected (Thompson \etal 1995; 
Fichtel \etal 1994). Their names, positions, redshifts, optical 
identifications and flux densities at 5~GHz are provided in Table~1.

\section{Observations}

Our VLBI observations were carried out within 48 hours on 1993 May
12$-$13, using the radio telescopes at Hartebeesthoek (South Africa), 
Hobart (Australia), Mopra (Australia), Parkes (Australia),
Perth (Australia), and Shanghai (China). The observing
parameters of these stations are given in Table~2. One element of
the Australia Telescope Compact Array (Narrabri, Australia) also observed,
but due to a configuration problem during recording, no useful
data were obtained.

This second snapshot session followed a similar observing mode and data
processing procedure as the first session (see Paper I), so only a brief
outline is given here. All 23 sources in Table~1 were observed
in snapshot mode, i.e., three to five 30-minute scans were obtained. 
Data were recorded in Mark~II format with 2-MHz bandwidth and
left-circular polarization (IEEE convention). The cross-correlation of the
data was carried out on the JPL/Caltech Mark~II Processor in 1994.

Post-correlation data reduction was done at the
Harvard-Smithsonian Center for Astrophysics, using the NRAO AIPS
and Caltech VLBI analysis packages.  A global fringe-fitting
procedure consisting of AIPS tasks was run at a solution interval of
2.5 minutes. The 64-m Parkes telescope was selected as the reference
station. Fringes were found for all 23 sources. The visibility data
were phase self-calibrated with a 10-second solution interval and a
point-source model in AIPS. The data averaging, editing, imaging, deconvolution,
and self-calibration were then performed within DIFMAP, a part of
the Caltech VLBI Package (Shepherd, Pearson \& Taylor 1994). Natural 
weighting was applied and only a constant gain factor correction was
implemented in the amplitude-calibration.
Finally, we ran the MODELFIT program in the Caltech VLBI Package to fit
the closure phases and amplitudes of the calibrated data on each
source, in order to obtain a quantitative description of its
structure. Up to three Gaussian components were fitted for 22 sources,
while 6 components were used in the case of 3C\,273, the strongest source in
the sample. In all cases, we feel that these models reasonably characterize
the fundamental features observed.

\section{Results}

The naturally weighted images for the 23 radio sources are presented
in Figure~1.  In each image, the size of the restoring beam is shown as
a cross-hatched ellipse in the lower-left corner. The lowest contour
level in each image is three times the rms noise level. The rms noise in the
images is a few mJys per beam for all the sources except 3C\,273, for which
the noise level was 18~mJy/beam.  The image parameters (peak flux
density, restoring beam and contour levels) are listed
in Table~3. The results of the model-fitting are given in Table~4,
together with the peak brightness
temperature for each model component in the rest frame of the source,
calculated following the method described in Paper I. 

\noindent
{\bf PKS~0118$-$272 (OC$-$230.4, Fig.~1A)}

\nobreak
This is a BL Lac object with a tentative emission-line redshift of
{\it z\/}=1.62 (\cite{ada85} and references therein). A lower limit of
{\it z\/}$>$~0.559 was suggested 
from several absorption lines (Falomo 1991; Stickel, Fried, \& K\"{u}hr 1993). 
Strong optical variations of 100\% within three days have been 
reported (Falomo, Scarpa, \& Bersanelli 1994). 
PKS~0118$-$272 is also highly optically polarized (Impey \& Tapia 1988;
Mead \etal 1990).

A 20-cm VLA observation revealed a complex structure, while at 6~cm
the source exhibited at least three components within
1$^{\prime\prime}$ (Perley 1982). A single-baseline 2.3-GHz VLBI
observation yielded a correlated flux density of 0.53$\pm$0.05,
corresponding to a visibility of 0.5$\pm$0.2 (Preston \etal 1985).
Our observations showed a resolved core with diffuse surrounding
emission.  The data were fitted by a single N--S elongated Gaussian
component with a flux density of 0.55~Jy. 
However, the core accounts for only half the total flux density of the source,
which supports the presence of two or three additional weak features in the
vicinity.
The source has a spectral index of 0.02 at millimeter
wavelengths (Steppe \etal 1988), with a total flux density of 0.60~Jy at
230~GHz.

\noindent
{\bf PKS~0208$-$512 (Fig.~1B)}

\nobreak
This flat-spectrum radio source has a redshift {\it z\/}=1.003 (Peterson \etal
1976).  It is classified as a highly polarized quasar or blazar by
Impey \& Tapia (1988, 1990). 
It was the second southern AGN (after 3C~279) detected at
$\gamma$-ray energies (\cite{brt93}; \cite{blo95}),
with one of the hardest photon spectral indices (von Montigny \etal 1995;
Chiang \etal 1995). 

No detailed arcsecond-scale structure of the source has been reported in the
literature. The SHEVE experiment at 2.3~GHz described it as a Gaussian 
component with a minimum nuclear flux density of 2.5~Jy (Preston \etal 1989). 
A 5-GHz VLBI image from 1992 November revealed a core-jet structure along a 
position angle of 233$^\circ$ (Tingay \etal 1996).

The image from our observation in 1993 May is in excellent agreement
with that of Tingay \etal (1996).  The source has a compact 2.8-Jy
core, and a 0.3-Jy jet component at an angular separation of 1.7~mas
and a position angle of 234$^\circ$ from the central core. A nearly flat
spectral index of 0.15 for the core is estimated from the 2.3- and
5.0-GHz results. The derived brightness temperature is
1.9~$\times~10^{12}$~K, close to the value of
1.2~$\times~10^{12}$~K in Tingay \etal (1996). We note that the
$\gamma$-ray emission is also thought to be beamed due to its short
variation time-scale (of tens of days) and high observed luminosity
(of 10$^{48}$~ergs~sec$^{-1}$ for isotropic emission) (\cite{brt93}).
Further study of this source may contribute to a better understanding of the
correlation between $\gamma$-ray emission and radio
radiation. A lower limit to the Doppler factor is 10.2 using the 
ROSAT X-ray observation (0.22~$\mu$Jy at 1~Kev) (Dondi \& Ghisellini
1995). This implies superluminal motion in the compact core, having
an upper limit of 6$^\circ$ to the viewing angle. Comparison of the
1992 and 1993 images taken 6.5 months apart suggests a proper motion of
0.6$\pm$0.7~mas~yr$^{-1}$,
corresponding to an apparent speed of 17$\pm$20~c.
Further high-accuracy VLBI measurements are needed to confirm its
superluminal motion.

\noindent
{\bf PKS~0332$-$403 (Fig.~1C)}

\nobreak
This source is a highly polarized quasar (Impey \& Tapia 1988; 1990). 
It has an inverted spectrum that peaks at around 5~GHz (Shimmins \etal
1971) and could be classified as a gigahertz-peaked-spectrum
(GPS) source (cf. O'Dea, Baum, \& Stanghellini 1991). 
Its redshift of {\it z\/}=1.445 has been widely
used in the literature without clear justification. 
This value is from {\sl Catalogue of Quasi Stellar-Objects}, edited
by Barbieri, Capaccioli, \& Zambon (1975), in which the reference was
incorrect. A new measurement should be made to confirm its redshift. 

The arcsecond structure of PKS~0332--403 is dominated by an unresolved
core from 6- and 20-cm VLA observations (Perley 1982). 
Preston \etal (1985) detected a correlated flux of 0.17$\pm$0.03~Jy,
in contrast to a total flux at 2.3~GHz of 4.0$\pm$0.4~Jy.
Single-baseline measurements at 2.3 and 8.4~GHz were also made
by Morabito \etal (1986). Our high-resolution image revealed a strong 
compact core and a weak unresolved feature to the east (see Table~4). 
The derived Doppler factor from the ROSAT 0.1$-$2.4~KeV observations
(Brinkmann, Siebert, \& Boller 1994) is greater than 3.6.  

\noindent
{\bf PKS~0403$-$132 (OF$-$105; Fig.~1D)}

\nobreak 
This quasar has strong emission lines and a redshift of {\it z\/}=0.571 (Lynds 1967). 
It is an optically violent variable source (Bolton \etal 
1966) and has a variable, modest linear polarization (Moore \& Stockman 1981;
Impey \& Tapia 1988). 
There is evidence of variability at X-ray energies (\cite{blu82}).

The VLA 20-cm image of Wardle \etal (1984) shows a core-jet structure, with an
unresolved nucleus extending on opposite sides, and a jet lying at a position
angle of 23$^\circ$. The nucleus remains unresolved at 6~cm (Morganti, Killeen,
\& Tadhunter 1993), with a possible weak extension (Wills \& Browne 1986). The
model from high-resolution VLBI data at 2.3~GHz exhibits two unresolved
components, which have a flux-density ratio of 5 and a separation of 580~mas
along a position angle of 48$^\circ$ (Preston \etal 1989).

Our VLBI image at 5~GHz revealed only a compact core, 
with no other component emission. The weak secondary component in
the 2.3-GHz model may be resolved, or have a steep spectrum, or have
decayed since 1982. The core component can be fitted by a single
0.5~mas~by~0.3~mas Gaussian component with a 0.85-Jy flux
density and a brightness temperature of 5.0~$\times$~10$^{11}$~K.
This high brightness is consistent with the
moderate variation of the source at radio (Romero, Benaglia, \& Combi 1995), 
optical (Bolton \etal 1966), and X-ray (\cite{blu82})
wavelengths. 

\noindent
{\bf PKS~0426$-$380 (Fig.~1E)}

\nobreak
This is a BL Lac object with no detected emission lines, and an absorption 
redshift of $z_{\small abs}$\/=1.030 (Stickel \etal 1993).
PKS~0426$-$380 has an optical polarization less than 3\% (Impey \& Tapia 1990). 
VLA observations at 20~cm show a 2.7-arcsecond extension from the
compact core to the position angle of $-$24$^\circ$ (Perley 1982). 
Single-baseline VLBI observations gave correlated flux densities of
0.70 and 0.90~Jy, at frequencies of 2.3 and 8.4~GHz, respectively
(Morabito \etal 1986).

The data of PKS~0426$-$380 were modeled by a Gaussian component 
extended along a position angle of 111$^\circ$.
Using the absorption redshift and the 
upper limit to the 1~keV X-ray flux density from ROSAT
observations (Brinkmann \etal 1994), the estimated lower
limit to the Doppler boosting factor is 3.1.  Our model
gives a brightness temperature of 7~$\times$~10$^{11}$~K for the
core.

\noindent
{\bf PKS~0438$-$436 (Figs.~1F \& 1F$^{\prime}$)}
 
\nobreak
This extremely luminous radio quasar has an emission-line
redshift of {\it z\/}=2.852 with some absorption lines superimposed (Morton,
Savage, \& Bolton 1978). Optically, PKS~0438$-$436 is faint
with variable polarization (e.g. Rusk 1990; Fugmann \& Meisenheimer 1988; 
Impey \& Tapia 1988, 1990).  It has a complex radio
spectrum, flat at low frequencies, but very steep above 5~GHz. 
The flux density at 1.4~GHz
is known to vary by 10\% over a period of four months.
The polarization also varies in degree and position angle (Luna \etal 1993).
It has been
detected by the IRAS satellite (Neugebauer \etal 1986), and also has
significant soft X-ray absorption (Wilkes \etal 1992). Both of these
properties are unusual for a quasar with such a high redshift. No
$\gamma$-ray detection was reported from the EGRET all-sky survey. 
Perley (1982) identified a secondary component 2.2~arcseconds away from the
core at a position angle of 15$^\circ$.  VLBI observations at 2.3~GHz
showed a core-jet structure: two 1.9~Jy circular components separated
by 35~mas at a position angle of $-$43$^\circ$ (Preston \etal 1989).

Our 5-GHz VLBI observation confirms the two-component structure described
above, with the same separation and orientation of the two components, to
within the errors. To facilitate a comparison, we convolved the 5.0-GHz data
with the 2.3-GHz SHEVE beam {(13.9~mas~$\times$~7.5~mas} at a position angle of
{58$^\circ$)} to obtain the image shown in {Fig.~F$^{\prime}$}, which has a peak
flux density of 1.6~Jy/beam. The lowest contour level is 23~mJy/beam, with a
factor of 2 between adjacent contours. It is clear that the northwest
component is stronger and more compact than the one to the southeast. We
identify the northwest component as the core, which has a brightness
temperature of {7.8~$\times$~10$^{11}$~K}, 100 times higher than that of the
other component. Actually, as can be seen in Fig.~1F, the southeast component
has been heavily resolved at 5~GHz. The core size agrees well with the analysis
of early VLBI measurement at 2.3~GHz, which showed a variable component
smaller than 1~mas in diameter (Gubbay \etal 1977). There is about a
{120$^\circ$} difference in position angle between the arcsecond- and
mas-scale structures.

\noindent
{\bf PKS~0454$-$234 (OF$-$292; Fig.~1G)}
 
\nobreak
This is listed as a BL Lac object (Ledden \& O'Dell 1985) due to its
featureless optical spectrum (Wilkes \etal 1983). An initial redshift
determination of 1.009 (Wright, Ables, \& Allen 1983) was later refined as 
{\it z\/}=1.003 (Stickel, Fried, \& K\"{u}hr 1989). It is a highly polarized
quasar with optical polarization up to 27\% (Wills \etal 1992). PKS~0454$-$234
was detected at marginal significance  by EGRET at $>$100~MeV 
$\gamma$-ray energies (Thompson \etal 1993b).

PKS~0454--234 was detected in the 2.3-GHz TDRSS experiment on a
1.8 earth-diameter baseline (Linfield \etal 1989), and in a 22-GHz
ground survey on a baseline of 10,000~km (Moellenbrock \etal 1996). 
Our 5-GHz VLBI image shows an asymmetrical morphology with a strong 
core and a compact jet-like component to the northwest at a position 
angle of $-62^\circ$.
The brightness temperature of the core as derived from our model is about 
6~$\times$~10$^{11}$~K. 
A Doppler beaming factor of 5.3 is estimated using the X-ray flux
density at 1~KeV and core structural parameters.  For comparison, a
value of 3.6 was derived from the variability time-scale (Dondi \&
Ghisellini 1995).

\noindent
{\bf PKS~0521$-$365 (Fig.~1H)}
 
\nobreak
This radio source has been identified with an N galaxy (Bolton \etal 1965).
A redshift of {\it z\/}=0.055  has been measured from both absorption features and
emission lines (Danziger \etal 1979). It is one of 5 radio sources
(cf. Crane \etal 1993) which have prominent optical counterparts to their
radio jets (Danziger \etal 1979; Folomo 1994 and references therein). 
The Hubble Space Telescope has resolved the optical jet
structure (Macchetto \etal 1991). High optical polarization has been
measured in the jet and the nucleus as well (Sparks, Miley, \& Macchetto 
1990 and references therein). This suggests a synchrotron
origin of the optical radiation. It has strong X-ray emission (Pian
\etal 1996) and has been marginally detected above 100~MeV
by EGRET (Fichtel \etal 1994; Lin \etal 1995).

Its arcsecond-scale radio structure is
dominated by an extended lobe to the southeast, rather than
by the compact core itself (Wardle, Moore, \& Angel 1984; Keel 1986; 
Slee \etal 1994). 
A radio jet follows closely, in direction
and extent, the optical jet to the northwest (Ekers \etal 1989). 
The magnetic fields inferred from the polarization measurements at the
optical and radio wavelengths are parallel to the jet direction in the jet, and
perpendicular to the jet in the core. From 2.3-GHz VLBI observations,
the core was modeled as a circular source 1.4 mas in diameter,
with a flux density of 1 Jy (Preston \etal 1989).

VLBI observations at 5.0 and 8.4~GHz revealed a 0.5-Jy jet component at
a position angle of 310$^\circ$ (Tingay \etal 1996). No apparent
motion could be determined from four-epoch measurements over a 
period of one year.

Our high-resolution data disclosed a second jet component near the compact core.
The parameters of these components are listed in Table~4. The two jet
components are aligned with the VLA jet (302$^\circ$$\pm$2$^\circ$, Keel 1986)
and optical jet ({$311^\circ\pm2^\circ$}, Cayatte \& Sol 1987). 
No beaming effect is needed for the core brightness temperature of 
1.7~$\times$~10$^{11}$~K. This is consistent with the nondetection of
superluminal motion. The $\gamma$-ray luminosity between 100~MeV and 5~GeV is
about {$3.2~\times~10^{44}$~ergs~sec$^{-1}$} (Lin \etal 1995), the second
lowest in the sample of gamma-ray-loud AGN. All of these observations suggest a
large angle in PKS~0521$-$365 between the ejection direction and the line of
sight. Pian \etal (1996) derived a viewing angle of 30$^\circ$ with bulk Lorentz
factor of 1.2. This leads to a predicted jet-counterjet ratio of 64, assuming a
jet spectral index of $-$1.0. However, none of the existing VLBI images reveal
any feature on the opposite side of the core. Further high dynamic range VLBI
images will help constrain the modelling of this source.

\noindent
{\bf PKS~0537$-$441 (Fig.~1I)}

\nobreak
This is a {\it z\/}=0.894 (Peterson \etal 1976) transition object between
classical BL Lac objects and quasars (Cristiani 1985; Maraschi \etal
1985; Giommi, Ansari, \& Micol 1995).  It has displayed substantial
variability at X-ray energies (Treves \etal 1993 and references
therein), and has shown similar variation time scales and amplitudes
at wavelengths ranging from infrared to X-ray (Tanzi \etal 1986),
suggesting that these emissions may originate from the same spatial
region.  The source is also a strong variable EGRET $\gamma$-ray
source (Thompson \etal 1993a). 

On the arcsecond scale the source is core-dominated with a bright 
secondary component separated by 7.2~arcseconds at a position angle of
305$^\circ$ (Perley 1982).
The SHEVE 2.3-GHz observation showed a 4.2-Jy 
core with a diameter of 1.1~mas (Preston \etal 1989). 
Recent VLBI observations at 4.9 and 8.4~GHz (Tingay \etal 1996) show
a jet-like component to the north of the compact core.

Our 5-GHz VLBI image confirmed the asymmetric core-jet structure,
in good agreement with the results from Tingay \etal (1996). However, the 
VLBI jet component differs by 70$^\circ$ in position angle from 
the VLA secondary component (Perley 1982).
The VLBI core has a brightness temperature of 8.6~$\times$~10$^{11}$~K.  
Unfortunately, the existing VLBI data were insufficient to determine the
proper motion of the jet in this EGRET-identified radio source.

\noindent
{\bf PKS~0823$-$223 (Fig.~1J)}

\nobreak
This is a BL Lac object (Wright \etal 1979; Wilkes \etal 1983)
with an absorption redshift of $z_{\small abs}$\/=0.910
(Falomo 1990 and references therein). 
It shows high optical polarization (Impey \& Tapia 1990). 
No {$\gamma$-ray} emission was detected by EGRET (Fichtel \etal 1994).

VLA observations show a diffuse structure on arcsecond scales at 20~cm (Perley
1982).  Our 6-cm VLBI image can be modeled by a {1.2~mas~by~0.3~mas}
Gaussian component with a flux density of 0.4~Jy.  The earlier
total flux-density measurement of 1.8~Jy (Impey \& Tapia 1990) and VLA
structure imply that PKS~0823$-$223 has some extended structure which is
resolved and therefore undetected by our VLBI observations.

\noindent
{\bf PKS~1034$-$293 (OL$-$259; Fig.~1K)}

\nobreak
This is a BL Lac object with a redshift of {\it z\/}=0.312 (Stickel \etal 1989 and references therein) and which displays high optical
polarization of up to 14\% (Wills \etal 1992).  PKS~1034$-$293 is a strong
millimeter-wavelength source with a variable flux density between
1.0~Jy to 3.0~Jy (Steppe \etal 1988, 1992, 1993).  

Previous VLBI observations described PKS~1034$-$293 as a core-dominated,
strong compact radio source (Robertson \etal 1993; Morabito
\etal 1986). The TDRSS experiment at 2.3~GHz fitted it as a
0.58-Jy, 0.44-mas circular Gaussian component (Linfield \etal 1989).

Our VLBI data were fitted with an elliptical Gaussian component
0.8~mas by 0.5~mas, with a flux density of 1.5~Jy. The derived
brightness temperature is 2.5~$\times$~10$^{11}$~K, consistent, within errors,
with the measurement of 4.6~$\times$~10$^{11}$~K from the 22-GHz ground
survey (Moellenbrock \etal 1996) and 9.2~$\times$~10$^{11}$~K from
the 2.3-GHz TDRSS (Linfield \etal 1989). 

\noindent
{\bf PKS~1127$-$145 (OM$-$146; Fig.~1L)}

\nobreak
This is a high-redshift quasar with broad emission lines at {\it z\/}=1.184
(Wilkes \etal 1983; Wilkes 1986) and absorption lines at $z_{\small
abs}$\/=0.313 (\cite{ber91}).  It has low optical
polarization (Impey \& Tapia 1990; Wills \etal 1992; Tornikoski \etal
1993) and is not an optically violently variable source (Pica \etal
1988; Bozyan, Hemenway, \& Argue 1990).  
It has been identified in the
second EGRET catalog at energies above 100~MeV (Thompson \etal 1995).

PKS~1127$-$145 was unresolved in VLA observations (Perley 1982). It is strong
and compact, and was detected at both 2.3 and 15~GHz on TDRSS baselines
larger than 14,000~km (Linfield \etal 1989, 1990). Two-epoch VLBI observations
at 1.7~GHz (Padrielli \etal 1986; Romney \etal 1984) found that PKS~1127$-$145 
was slightly extended to the north. No structural variation between the two
epochs (1.7~years) was seen. A proper motion of 0$\pm$0.02 in the source was
reported by Vermeulen \& Cohen (1994). The 5-GHz image of Wehrle \etal (1992)
resolved the source into two compact components plus a weak extension to the
northeast.

Our VLBI observation revealed two Gaussian components. The brighter component,
with a flux density of 1.9~Jy, is larger than the weaker (1.1~Jy) component
(see Table~4). These two components have similar brightness temperatures of
{$\sim$1.3~$\times$~$10^{11}$~K}. They are probably related to two components of
nearly equal strength resolved by Wehrle \etal (1992), and possibly to the
east--west structure observed at 1.7~GHz (Padrielli \etal 1986; Romney \etal
1984). We find no evidence of a northeast extension seen by Wehrle \etal
(1992). We note that early VLBI experiments at 18, 13 and
6~cm indicated at least three distinct components in the source,
implying source evolution (Kellermann \etal 1971; Weiler \& de Pater 1983). 

\noindent
{\bf PKS~1226+023 (3C~273; Fig.~1M)}

\nobreak
This well-known radio source has been identified with a thirteenth magnitude
object at a redshift of {\it z\/}=0.158 (Hazard, Mackey, \& Shimmins 1963; Schmidt
1963). It shows optical variation, which was first analyzed by Smith and
Hoffleit (1963), but does not have high optical polarization (Appenzeller 1968).
It is very bright across the wavebands from radio to {$\gamma$-ray}. It was the
only {$\gamma$-ray} source known before the CGRO observations (Swanenburg \etal
1978; \cite{big81}), and one of the first two extragalactic sources detected by
EGRET (Mattox \etal 1997 and references therein).

The large-scale structure from the VLA and MERLIN shows a compact,
flat-spectrum core and a single jet extending about 23 arcseconds from
the core at a position angle of 222$^\circ$ (Conway \etal 1993).
This source has received considerable attention since the VLBI
technique became available, due mainly to its intensity and variability. 
Multi-frequency VLBI observations show a bright core 
and a number of jet components extending toward the
southwest (e.g. Davis \etal 1991; Zensus \etal 1988). 

Our VLBI observation of  this equatorial quasar 
has a good north--south resolution with a beam
of 1.6~mas by 0.88~mas at a position angle of 33$^\circ$. 
We fitted six components, labeled 1 through 6, to the data.
The strong component 1 at the eastern end is identified as the core. 
No counterjet is visible. Components 2 to 6 are jet components, 
or knots in the continuous jet, which have a similar position angle 
of {$\sim230^\circ$} and increasing distance to the core (from 1.9~mas 
for component 2 to 15.2~mas for component 6). Along this position angle,
there is a distinct emission gap between components 5 and 6. Such
morphology is consistent with other published results (e.g. Zensus \etal 
1988). Comparison with earlier observations enables us to identify
the components in our image with those seen previously:
our components 2, 3 and 4 are respectively C$_{\small 10}$, C$_{\small 9}$ and
C$_{\small 8}$ (\cite{abr94}).
Our component 5 is C$_{\small 7a}$ (Cohen \etal 1987).
The more extended component 6 in our image is more difficult to identify,
and may be C$_{\small 6}$ (Unwin \etal 1985; Zensus 1987; Charlot, Lestrade,
\& Boucher 1988) or possibly a mixture of C$_{\small 6}$ and other components
(such as C$_{\small 5}$, C$_{\small 4}$, or even C$_{\small 3}$)   
(see Cohen \etal 1987, Unwin \etal 1985). Such identification is in good
agreement with the observational picture of the evolution of the different
components in 3C~273 (see \cite{abr96}).

\noindent
{\bf PKS~1244$-$255 (Fig.~1N)}

\nobreak
This source is a blazar (\cite{brs92}), having violent
optical variation (Bozyan \etal 1990 and references
therein) and high optical polarization (Impey \& Tapia 1988, 1990). It
has strong emission lines corresponding to a redshift of {\it z\/}=0.638 (Falomo \etal
1994).

VLBI observations yielded correlated flux densities of 0.48~Jy at 2.3~GHz,
1.01~Jy at 8.4~GHz (Morabito \etal 1986), and 0.91~Jy at 22~GHz 
(Moellenbrock \etal 1996). Our VLBI image
shows a simple compact core elongated at a position angle of
116$^\circ$, with a brightness temperature of 5.6~$\times$~10$^{11}$~K.
This is consistent with the measurement of $>$4.0~$\times$~10$^{11}$~K from
the 22-GHz survey. 

\noindent
{\bf PKS~1424$-$418 (Fig.~1O)}

\nobreak
This is a highly optically polarized quasar (Impey \& Tapia 1988,
1990). An accurate redshift measurement of {\it z\/}=1.524 was made by Stickel
\etal (1989). 
Data on PKS~1424$-$418 from the 2.3-GHz SHEVE observations
were modeled by two circular components separated by 23~mas
(Preston \etal 1989). Our VLBI
image shows two components separated by $\sim$3~mas.
Assuming the stronger component is the core, the  
position angle of the weaker component is 260$^\circ$, significantly
different from the reported value for the two 2.3-GHz components of
236/284$^\circ$ (Preston \etal 1989). 
There is a difference in alignment of 90$^\circ$ between the VLBI and VLA
structures.
Using the model results, we obtain a very flat spectral index of
$-$0.04 for the central core. For comparison, we calculated a spectral
index of 0.20 from the correlated flux density measured at 2.3 and
8.4~GHz (Robertson \etal 1993). An asymmetry was also inferred from
their data. 

\noindent
{\bf PKS~1514$-$241 (AP LIB; Fig.~1P)}

\nobreak
This source is a classical BL Lac object (Strittmatter \etal 1972).
The redshift is {\it z\/}=0.0486, 
based upon both absorption lines and emission lines (Rodgers \&
Peterson 1977 and references therein). It has been characterized as an optically
violent variable (Carini \etal 1991; Bozyan \etal 1990;
Webb \etal 1988) and a highly polarized quasar (Wills \etal 1992). 

This source shows a core-jet morphology on arcsecond scales in 6- and
20-cm VLA images (Morganti \etal 1993; \cite{ant85}; 
Ulvestad, Johnston, \& Weiler 1983). 
A component 0.2~arcsecond from the core at a position angle of 120$^\circ$ 
was reported by Perley (1982). 

Our VLBI image of this BL Lac object shows only a single component, 1.2~mas
by 0.6~mas in size, with elongation in the north--south direction, and a
flux density of 1.53~Jy.
The brightness temperature is 1.1~$\times$~10$^{11}$~K, 
which is consistent with the value of $\sim$1.5~$\times$~10$^{11}$~K 
from the 22-GHz survey (Moellenbrock \etal 1996). 

\noindent
{\bf PKS~1936$-$155 (OV$-$161; Fig.~1Q)}

\nobreak
This is a blazar with high optical polarization (Fugmann \&
Meisenheimer 1988) and a high redshift of {\it z\/}=1.657 (Jauncey \etal
1984). It has a very steep spectrum between 1.4 and 2.7~GHz, and a flat
spectrum between 2.7 and 5.0~GHz, and probably to 8.4~GHz (Quiniento
\& Cersosimo 1993; Impey \& Tapia 1990). 

VLA 6-cm observations showed the source to be 
unresolved, with a flux density of 0.72~Jy (Neff, Hutchings, \& Gower 1989). 
Our VLBI image shows a single component 0.6~mas by 0.5~mas in size,
elongated in the north--south 
direction, with a flux density of 0.97~Jy. 
A slight extension to the southeast can be seen from the map.
The compact core has a brightness temperature of 4.7~$\times$~10$^{11}$~K,
compared to $>$2.5~$\times$~10$^{11}$~K from the 22-GHz survey (Moellenbrock 
\etal 1996). 

\noindent
{\bf PKS~1954$-$388 (Fig.~1R)}

\nobreak
This is an optically violent variable (Gilmore 1980) with a redshift of
{\it z\/}=0.626 (Browne, Savage, \& Bolton 1975). It also has a high optical
polarization up to 11\% (Impey \& Tapia 1988, 1990). 
VLA observations at
20~cm showed a diffuse structure (Perley 1982). Previous VLBI observations
showed the presence of a compact
core at 2.3~GHz (Preston \etal 1985).
Our VLBI image shows a single compact component 0.4~mas by 0.3~mas in size,
with a flux density of 1.9~Jy.
The calculated brightness temperature is 1.2~$\times$~10$^{12}$~K.

\noindent
{\bf PKS~2005$-$489 (Fig.~1S)}

\nobreak
This source is a BL Lac object at a redshift of {\it z\/}=0.071 (Falomo \etal 1987). 
It is one of two objects with low optical polarization in the 1-Jy BL Lac
object sample (Stickel \etal 1993; Stickel \etal 1991). 
It is one of the few extragalactic sources detected in the EUV band 
(Marshall, Fruscione, \& Carone 1995). Its
detected X-ray flux is far in excess of a simple power-law
extrapolation from lower frequency measurements, and is among the three or
four brightest BL Lac objects having violent optical variation 
(Wall \etal 1986; Bozyan \etal 1990). 
It has shown large amplitude, short time-scale
variability among different X-ray observations (e.g. Della Ceca \etal 1990; 
Elvis \etal 1992; Brinkmann \etal 1994; Ghosh \& 
Soundararajperumal 1995). It was only marginally detected by EGRET
(Fichtel \etal 1994; Thompson \etal 1995).

Our VLBI data shows a simple compact component, {1.0~mas by 0.2~mas} in size,
with a flux density of 0.92~Jy and a corresponding brightness temperature of
{$2.6~\times~10^{11}$~K}.  

\noindent
{\bf PKS~2155$-$152 (OX$-$192; Fig.~1T)}

\nobreak
This is an OVV BL Lac object (Craine \etal 1976; Bozyan \etal 1990)
displaying very high optical polarization (Brindle \etal 1986; Impey 
\& Tapia 1990) and a redshift of {\it z\/}=0.672 (White \etal 1988; 
Stickel \etal 1989). 
On the arcsecond scale, the source exhibits a triple structure with an
extension of 6~arcseconds along the north--south direction (Weiler \&
Johnston 1980; Perley 1982; Wardle \etal 1984).
Our VLBI image reveals a core-jet morphology. 
The unresolved core is strong, with a high brightness 
temperature of $>$6.7~$\times$~10$^{11}$~K.
The source axis seems to be well aligned on both the mas and 
arcsecond scales. 

\noindent
{\bf PKS~2240$-$260 (OY$-$268; Fig.~1U)}

\nobreak
This BL Lac object has a redshift of {\it z\/}=0.774 (Stickel \etal 1993). 
High optical polarization has also been observed in the source (Impey \& Tapia
1988, 1990). Our VLBI data shows a core-jet structure, of which the stronger
component has a flux density of 0.54~Jy. The secondary component, at a
distance of 2.2~mas and a position angle of {$314^\circ$}, is 0.12~Jy. We
identify the core with the stronger component, although both components have
similar brightness temperatures.

\noindent
{\bf PKS~2243$-$123 (OY$-$172.6; Fig.~1V)}

\nobreak
This is an optically violent variable and a highly polarized quasar
(Impey \& Tapia 1988, 1990; Wills \etal 1992), with a redshift of
{\it z\/}=0.630 (Browne \etal 1975). This radio source has also 
been classified as a GPS source, because its spectrum shows a turnover
at $\sim$2.3~GHz (Cersosimo \etal 1994).  

VLA observations exhibited an unresolved core with a 4-arcsecond
extended component at a position angle of 40$^\circ$ (Perley 1982;
Browne \& Perley 1986; Morganti \etal 1993). 
Observations on a single long baseline measured correlated flux densities
of 0.78~Jy and 1.23~Jy, at 2.3 and 8.4~GHz, respectively (Morabito
\etal 1986).

Our VLBI data can be fitted as a compact core with a flux 
density of 2.28~Jy, and size of 1.1~mas by 0.4~mas, with 
a corresponding brightness temperature of 4.0~$\times$~10$^{11}$~K.
The core has a north--south elongation. We note that there is a
depression at the east--west side of the core, which might indicate
the emergence of a new component. 
A second epoch VLBI observation in 1995 October
showed a resolved structure, with a jet-like component to the south,
1.12~mas from the strong central component (Shen, Hong, \& Wan 1998). 
A proper motion of {0.22~mas~yr$^{-1}$} is estimated, which corresponds to an
apparent superluminal motion of 4.6c in the jet. 

\noindent
{\bf PKS~2355$-$534 (Fig.~1W)}

\nobreak
This is an optically violent and highly polarized source (Impey \& Tapia 1988,
1990) with a high redshift of {\it z\/}=1.006 (Jauncey \etal 1984).
There have been no previous measurements of the radio structure.
Our image shows two components with similar size but different flux 
densities. The stronger component, which may be the core,  has a brightness
temperature of 3.5~$\times$~10$^{11}$~K. The second component is
located at a distance of 4.9~mas at a position angle of 235$^\circ$. 

\section{Summary}
 
In this paper we have defined a sample of southern-hemisphere core-dominated
blazars.  Of the 24 blazars in the sample, 3 were observed earlier with the
same array. The other 21 in the sample and 2 other sources were observed
in 1993 May with the Southern VLBI Network plus the Shanghai radio telescope. 
This is part of the Southern Hemisphere 5-GHz VLBI Survey project, the aim of
which is to improve the study of southern extragalactic radio sources (see
Paper I). Our study also adds significantly to the number of sources
whose structures can be compared on arcsecond (kpc) and milliarcsecond (pc)
scales (Table~5). The misalignment of jet-like structures on these scales is
an important unsolved problem for the understanding of compact sources.

The main conclusions presented in this paper can be summarized as follows:

\begin{enumerate}
\item
We have detected and imaged all 23 radio sources, of which
15 are first-epoch VLBI images. These are 
PKS~0118$-$272, PKS~0332$-$403, PKS~0426$-$380, PKS~0454$-$234, 
PKS~0823$-$223, PKS~1034$-$293, PKS~1244$-$255, PKS~1514$-$241, 
PKS~1936$-$155, PKS~1954$-$388, PKS~2005$-$489, PKS~2155$-$152, 
PKS~2240$-$260, PKS~2243$-$123 and PKS~2355$-$534. 
\item
Most of the blazars are resolved and display simple morphology, with
12 having core-jet structures and 11 having single-core
structures. Observations
with increased sensitivity will probably reveal many more
core-jet structures (e.g. 2243--123).
We have compared our VLBI images with other radio images.
Only 3 (PKS~0438$-$436, PKS~0537$-$441 and PKS~1226+023) of the
12 core-jet blazars were found to have curved jets.  
Superluminal motion was inferred from two-epoch observations for 2
sources (PKS~0208$-$512, PKS~2243$-$123). 
\item
Eight of these blazars (PKS~0208$-$512, PKS~0454$-$234, PKS~0521$-$365, 
PKS~0537$-$441, PKS~1127$-$145, PKS~1226+023, PKS~1424$-$418 and 
PKS~2005$-$489) have been detected at $>100$~MeV $\gamma$-ray energies. 
Together with the other 5 EGRET sources observed in 1992 November
(Paper I), a total of 13 southern $\gamma$-ray-loud
blazars have now been imaged by our survey project. A systematic study
of the VLBI properties of these $\gamma$-ray blazars and comparison with
other non-$\gamma$-ray sources will improve our understanding of
the beaming characteristics in blazars and the properties of EGRET
sources.
\end{enumerate}

\acknowledgments
This work was supported at Shanghai Astronomical Observatory by grants from
the National Program for the Enhancement of Fundamental Research.
Part of this research was carried out at the Jet Propulsion
Laboratory, California Institute of Technology, under
contract with the National Aeronautics and Space Administration.
We would like to thank M. Reid, M. Birkinshaw and C. Carilli
for helpful discussions. We thank an anonymous referee for helpful and
constructive comments.
Z.-Q. Shen acknowledges the receipt of a Smithsonian Pre-doctoral Fellowship.
The Australia Telescope National Facility is funded by the Australian 
Government for operation as a national facility by the CSIRO.
Our search of the literature was greatly assisted by 
the NASA/IPAC Extragalactic Database~(NED), which is operated by 
the Jet Propulsion Laboratory, California Institute of Technology, 
under contract with the National Aeronautics and Space Administration.

\clearpage

{\setlength{\baselineskip}{13pt}

\clearpage

{\bf \large FIGURE CAPTION}

\noindent
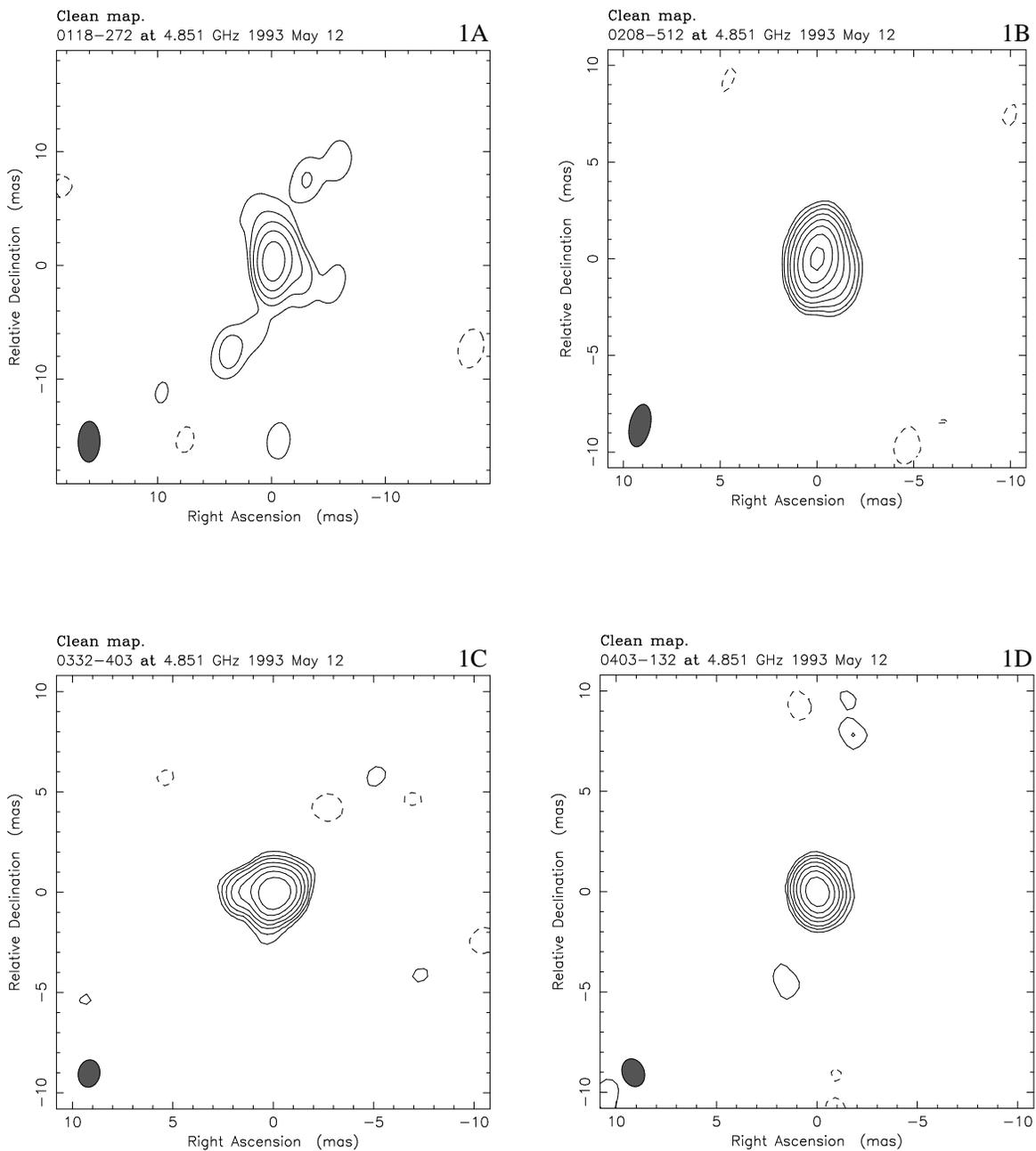
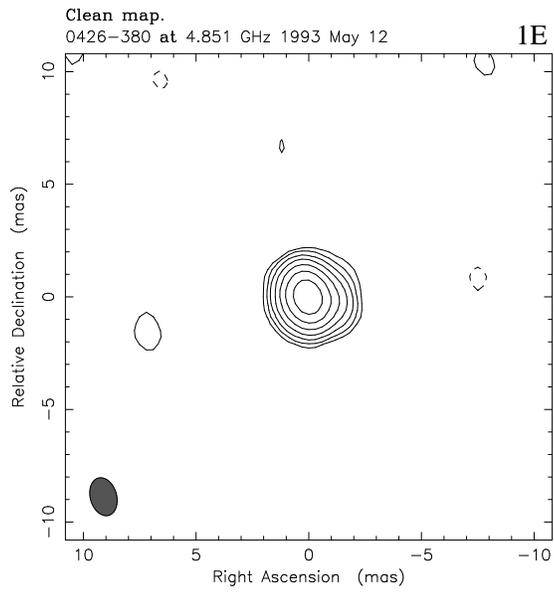
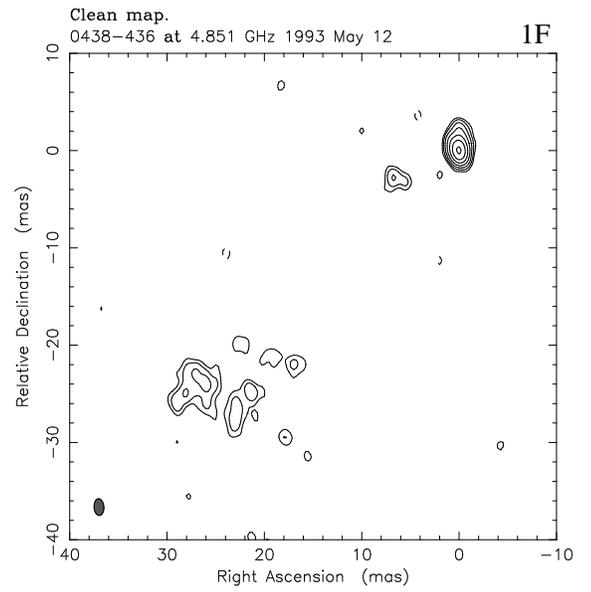
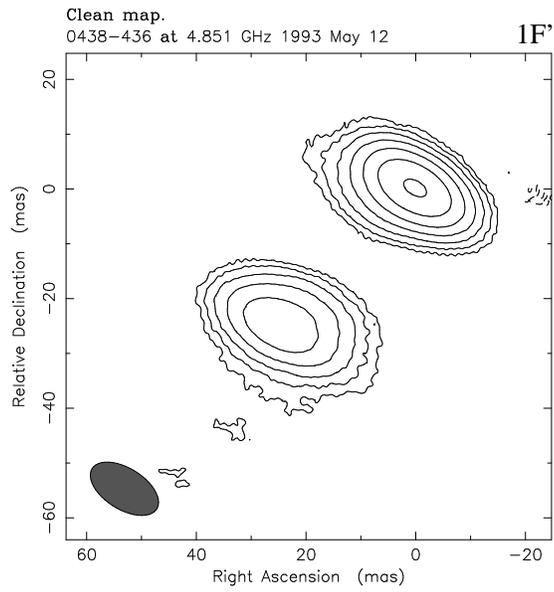
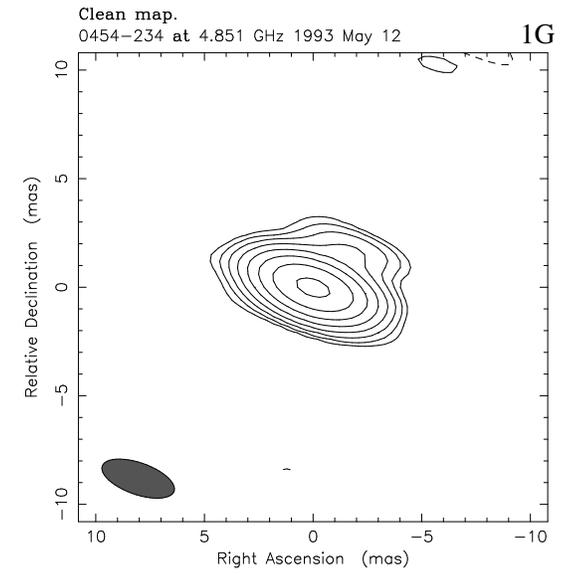
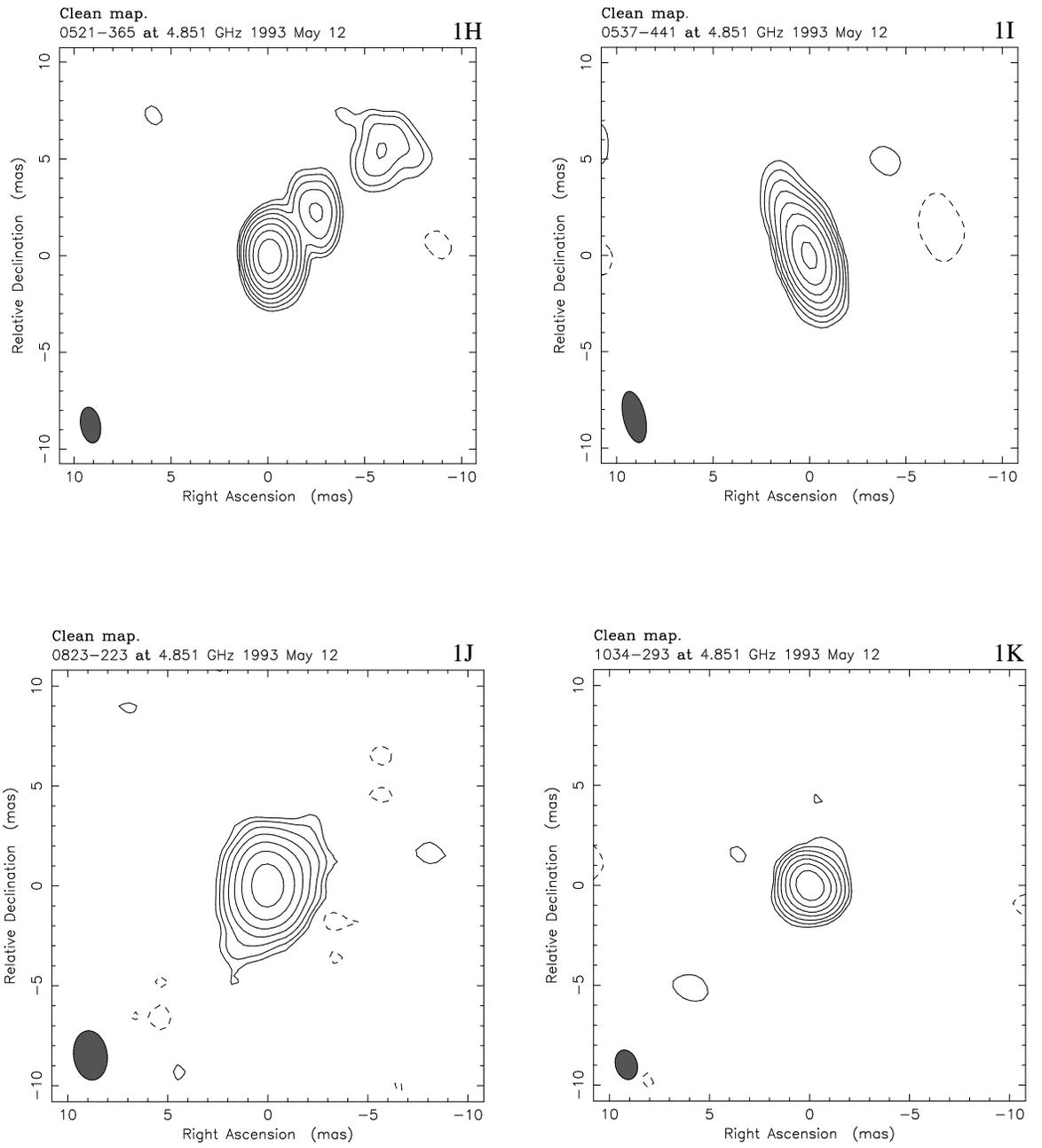
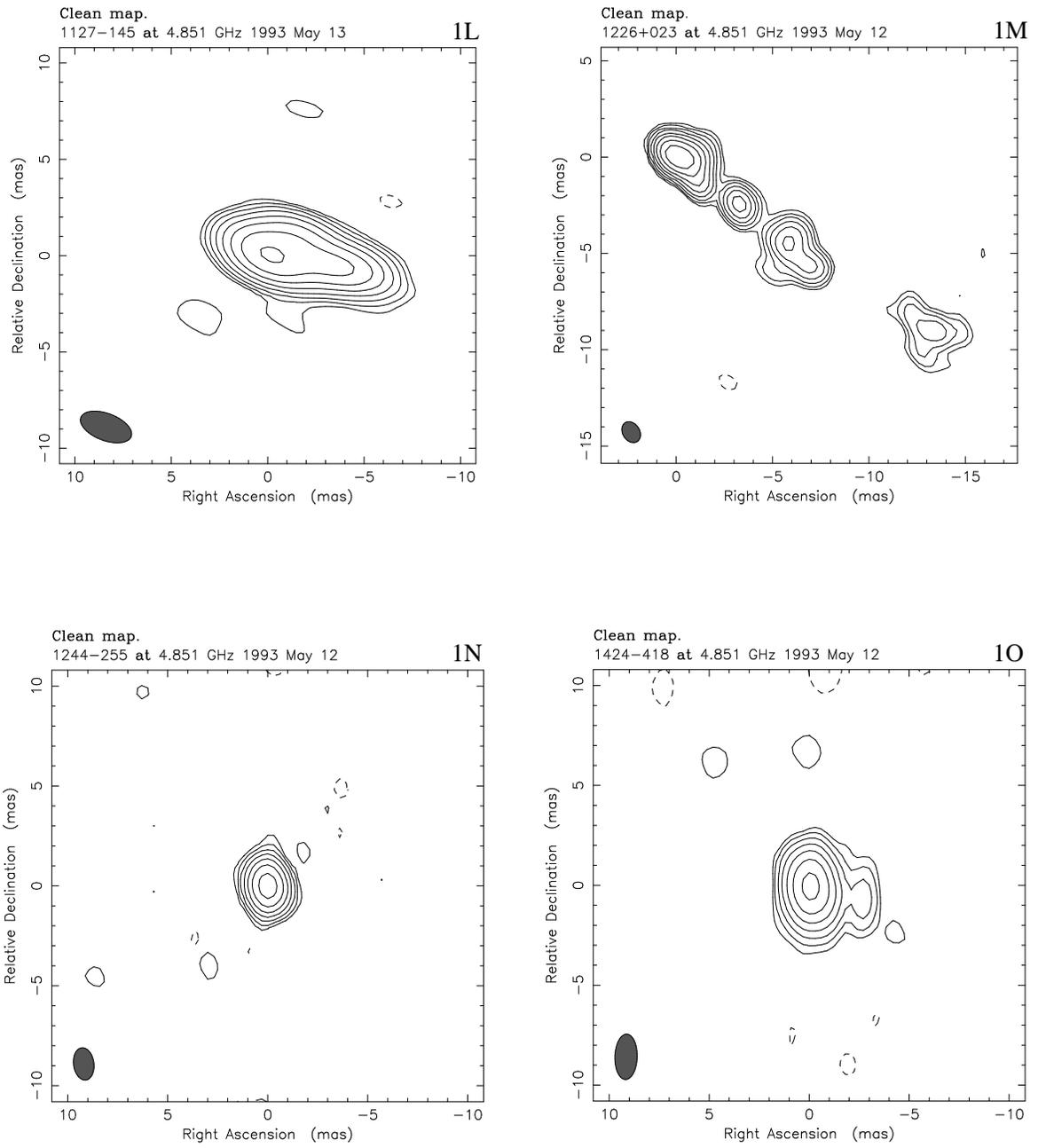
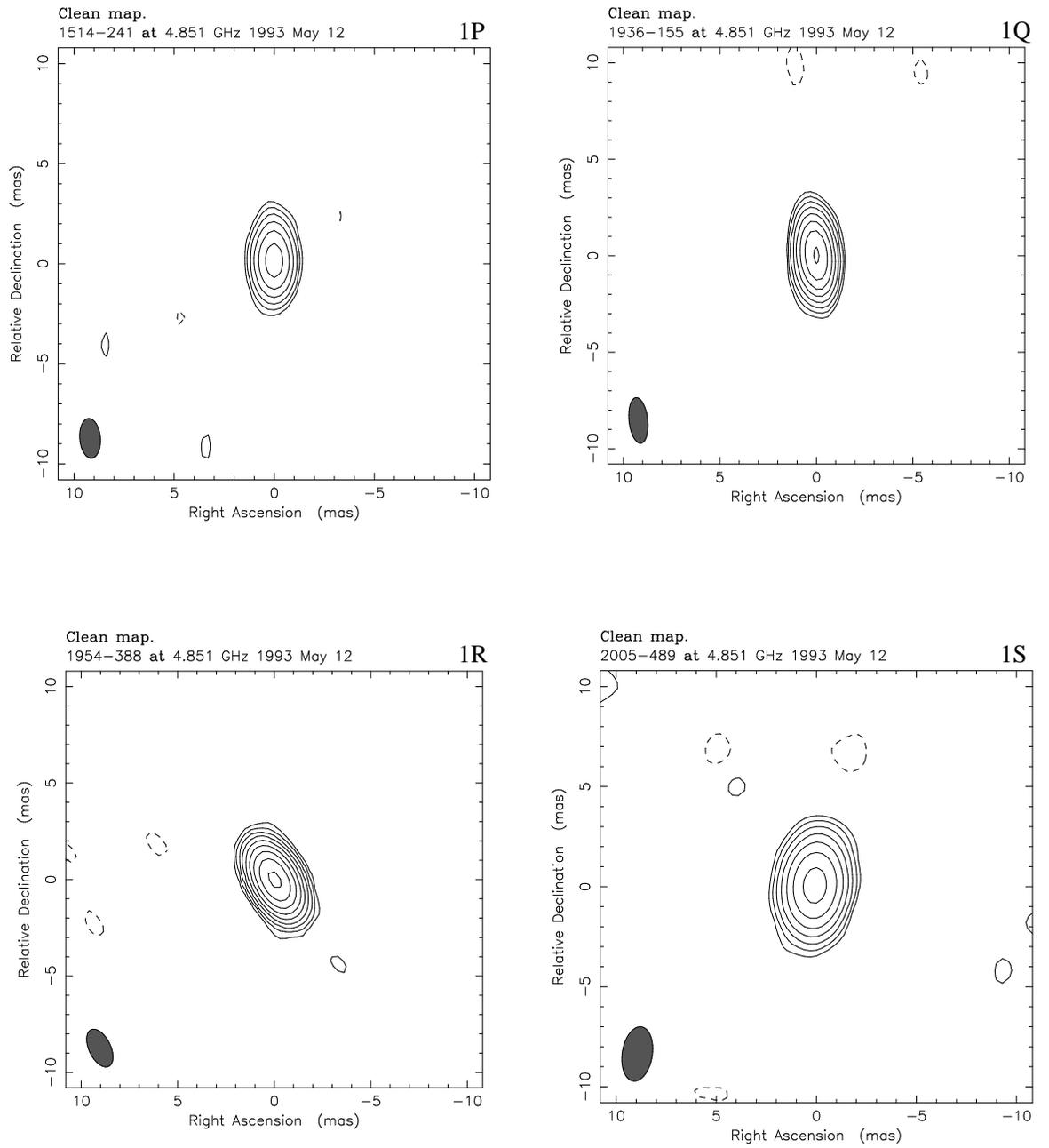
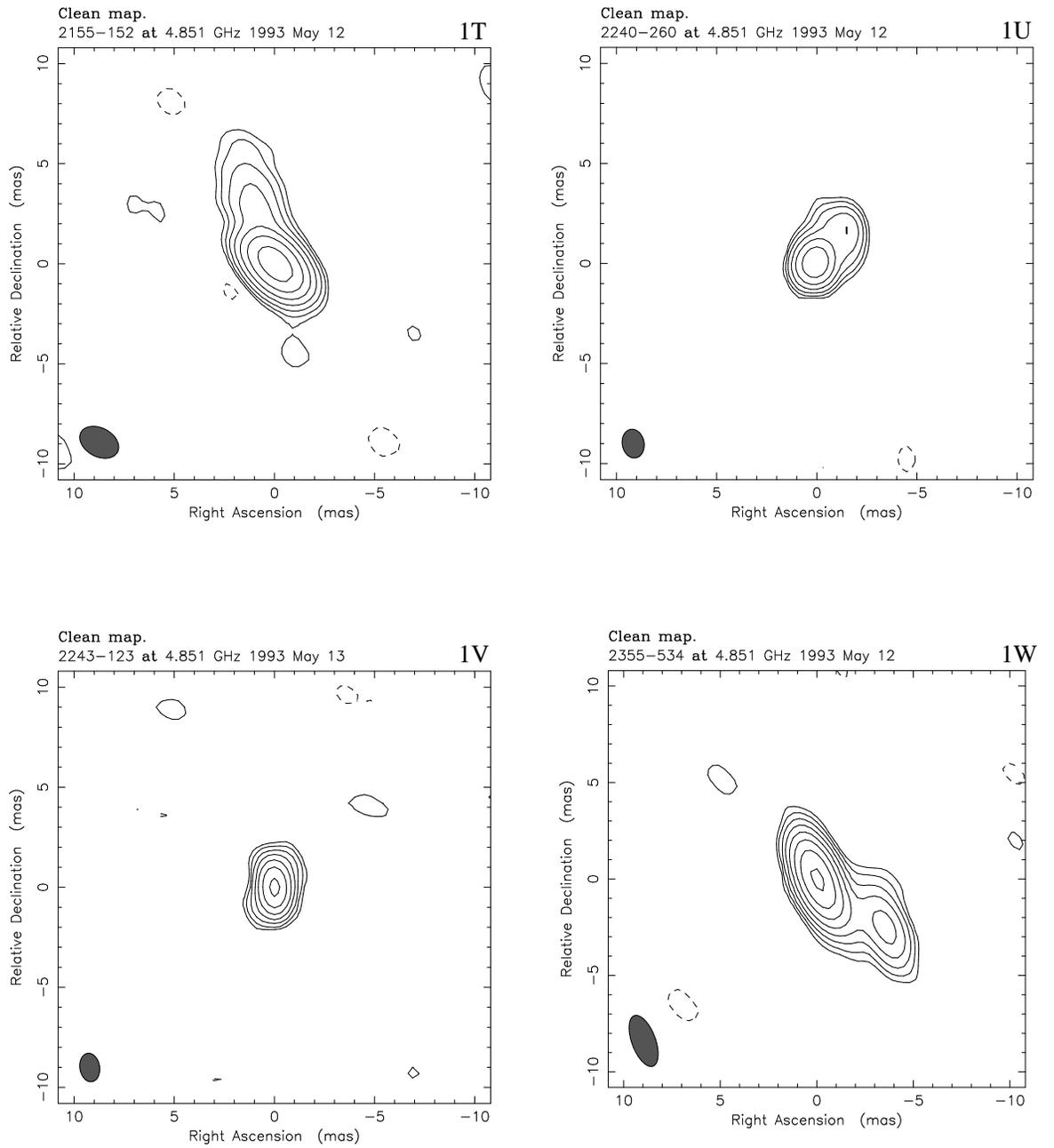
{\bf Fig. 1}--- VLBI images of the 23 extragalactic radio sources observed 
in May 1993. The synthesized beam is shown in the lower left of each image. 
See Table~3 for detailed imaging parameters.


\begin{thebibliography}{}

\bibitem[Abraham \etal 1994]{abr94}
	Abraham, Z., Carrara, E.~A., Zensus, J.~A., \& Unwin, S.~C. 1994,
	in Compact Extragalactic Radio Sources, eds. J.~A. Zensus and
	K.~I. Kellermann (National Radio Astronomy Observatory, Greenbank), 87-90 
\bibitem[Abraham \etal 1996]{abr96}
	Abraham, Z., Carrara, E.~A., Zensus, J.~A., \& Unwin, S.~C.
	1996, \aaps, 115, 543-549
\bibitem[Adam 1985]{ada85}
	Adam, G.  1985, \aaps, 61, 225-235
\bibitem[Antonucci \& Ulvestad 1985]{ant85}
	Antonucci, R.~R.~J., \& Ulvestad, J.~S. 1985, \apj, 294, 158-182
\bibitem[Appenzeller 1968]{apz68}
	Appenzeller, I. 1968, \apj, 151, 769-770
\bibitem[Barbieri, Capaccioli, \& Zambon 1975]{bar75}
	Barbieri, C., Capaccioli, M., \& Zambon, M. 1975, Mem. Soc. Astron.
	Italiana, 46, 461-499
\bibitem[Bergeron \& Boisse 1991]{ber91}
	Bergeron, J., \& Boisse, P. 1991, \aap, 243, 344-366
\bibitem[Bersanelli \etal 1992]{brs92}
	Bersanelli, M., Bouchet, P., Falomo, R., \& Tanzi, E. G. 1992, \aj, 104, 28-39
\bibitem[Bignami \etal 1981]{big81}
	Bignami, G. F., \etal 1981, \aap, 93, 71-75
\bibitem[Bertsch \etal 1993]{brt93}
	Bertsch, D. L., \etal 1993, \apj, 405, L21-L24
\bibitem[Blom \etal 1995]{blo95}
	Blom, J. J.,  \etal 1995, \aap, 298, L33-L36
\bibitem[Blumenthal, Keel, \& Miller 1982]{blu82}
	Blumenthal, G. R., Keel, W. C., \& Miller, J. S. 1982, \apj, 257, 499-508
\bibitem{}Bolton, J. G., Clarke, M. E., \& Ekers, R. D. 1965, Australian J. Phys., 18, 627
\bibitem{}Bolton, J. G., Savage, A., \& Wright, A. E. 1979, Aust. J. Phys. Astroph. Supp, 46, 1-17
\bibitem{}Bolton, J. G., Shimmins, A. J., Ekers, J., Kinman, T. D., Lamia, E.,
	\& Wirtanen, C.~A. 1966, \apj, 144, 1229-1231
\bibitem{}Bozyan, E. P., Hemenway, P. D., \& Argue, A. N. 1990, \aj, 99, 1421-1434
\bibitem{}Brindle, C., Hough, J. H., Bailey, J. A., Axon, D. J., \& Hyland, A. R. 1986, \mnras, 221, 739-768
\bibitem{}Brinkmann, M., Siebert, J., \& Boller, T. 1994, \aap, 281, 355-374
\bibitem{}Browne, I. W. A., Savage, A., \& Bolton, J. G. 1975, \mnras, 173, 87
\bibitem{}Browne, I. W. A., \& Perley, R. A. 1986, \mnras, 222, 149-166
\bibitem{}Burbidge, G., \& Hewitt, A. 1992, in Variability of Blazars, eds. E. Valtaoja and M. Valtonen
	(Cambridge University Press, Cambridge), 4-38
\bibitem{}Carini, M. T., Miller, H. R., Noble, J. C., \& Sadun, A. C. 1991, \aj, 101, 1196-1201
\bibitem{}Cayatte, V., \& Sol, H. 1987, \aap, 171, 25-32
\bibitem{}Cersosimo, J. C., Lebron Santos, M., Cintron, S. I., \& Quiniento, Z. M. 1994, \apjs, 95, 157-161
\bibitem{}Charlot, P., Lestrade, J.-F., \& Boucher, C. 1988, in IAU Symp.~129, The Impact of VLBI
        on Astrophysics and Geophysics, eds. M.~J. Reid and J.~M. Moran (Kluwer, Dordrecht), 33-34
\bibitem{}Chiang, J., \etal 1995, \apj, 452, 156-167
\bibitem{}Cohen, M. H., Zensus, J. A., Biretta, J. A., Comoretto, G., Kaufmann, P., \& Abraham, Z. 1987
	\apjl, 315, L89-L92
\bibitem{}Conway, R. G., Garrington, S. T., Perley, R. A., \& Biretta, J. A. 1993, \aap, 267, 347-362
\bibitem{}Craine, E. R., \etal 1976, \aplett, 17, 123-125
\bibitem{}Crane, P., \etal 1993, \apjl, 402, L37-L40
\bibitem{}Cristiani, S. 1985, IAU Circulars, No. 4027
\bibitem{}Danziger, I. J., Fosbury, R. A. E., Goss, W. M., \& Ekers, R. D. 1979, \mnras, 188, 415-419
\bibitem{}Davis, R. J., Unwin, S. C., \& Muxlow, T. W. B. 1991, \nat, 354, 374-376
\bibitem{}Della Ceca, R., Palumbo, G. G. C., Persic, M., Boldt, E. A., de Zotti, G.,
	\& Marshall, E.~E. 1990, \apjs, 72, 471-550
\bibitem{}Dondi, L., \& Ghisellini, G. 1995, \mnras, 273, 583-593
\bibitem{}Ekers, R. D., \etal 1989, \mnras, 236, 737-777
\bibitem{}Elvis, M., Plummer, D., Schachter, J., \& Fabbiano, G. 1992, \apjs, 80, 257-303
\bibitem{}Falomo, R., Maraschi, L., Tanzi, E. G., \& Treves, A. 1987, \apjl, 318, L39-L41
\bibitem{}Falomo, R. 1990, \apj, 353, 114-117
\bibitem{}Falomo, R. 1991, \aj, 102, 1991-1993
\bibitem{}Falomo, R. 1994, ESO Messenger, 77, 49-52
\bibitem{}Falomo, R., Scarpa, R., \& Bersanelli, M. 1994, \apjs, 93, 125-143
\bibitem{}Fichtel, C. E., \etal 1994, \apjs, 94, 551-581
\bibitem{}Fugmann, W., \& Meisenheimer, K. 1988, \aaps, 76, 145-156
\bibitem{}Ghosh, K. K., \& Soundararajperumal, S. 1995, \apjs, 100, 37-68
\bibitem{}Gilmore, G. 1980, \mnras, 190, 649-667
\bibitem{}Giommi, P., Ansari, S. G., \& Micol, A. 1995, \aaps, 109, 267-291
\bibitem{}Gubbay, J., Legg, A. J., Robertson, D. S., Nicolson, G. D., Moffet, A. T. \&
	Shaffer, D. B. 1977, \apj, 215, 20-35
\bibitem{}Hazard, C., Mackey, M. B., \& Shimmins, A. J. 1963, \nat, 197, 1040
\bibitem{}Henstock, D. R., \etal 1995, \apjs, 100, 1-36
\bibitem{}Impey, C. D., \& Tapia, S. 1988, \apj, 333, 666-672
\bibitem{}Impey, C. D., \& Tapia, S. 1990, \apj, 354, 124-139
\bibitem{}Impey, C. D. 1992, in Variability of Blazars, eds. E. Valtaoja and M. Valtonen
	(Cambridge University Press, Cambridge), 55-69
\bibitem{}Jauncey, D. L., Batty, M. J., Wright, A. E., Peterson, B. A., \& Savage, A. 1984,
	\apj, 286, 498-502
\bibitem{}Keel, W. C. 1986, \apj, 302, 296-305
\bibitem{}Kellermann, K. I., \etal 1971, \apj, 169, 1-25
\bibitem{}Ledden, J. E., \& O'Dell, S. L. 1985, \apj, 298, 630-643
\bibitem{}Linfield, R. P., \etal 1989, \apj, 336, 1105-1112
\bibitem{}Linfield, R. P., \etal 1990, \apj, 358, 350-358
\bibitem{}Lin, Y. C., \etal 1995, \apj, 442, 96-104
\bibitem{}Luna, H. G., Martinez, R. E., Combi, J. A., \& Romero, G. E. 1993, \aap, 269, 77-82
\bibitem{}Lynds, C. R.  1967, \apj, 147, 837-840
\bibitem{}Macchetto, F., \etal 1991, \apj, 369, L55-L57
\bibitem{}Maraschi, L., Schwartz, D. A., Tanzi, E. G., \& Treves, A., 1985, \apj, 294, 615-618
\bibitem{}Marshall, H. L., Fruscione, A., \& Carone, T. E. 1995, \apj, 439, 90-97
\bibitem{}Mattox, J. R., Schachter, J., Molnar, L., Hartman, R. C., \& Patnaik, A. R. 1997, \apj, 481, 95-115
\bibitem{}Mead, A. R. G., Ballard, K. R., Brand, P.~W.~J.~L., Hough, J.~H.
	Brindle, C. \& Bailey, J.~A. 1990, \aaps, 83, 183-204
\bibitem{}Moellenbrock, G. A., \etal 1996, \aj, 111, 2174
\bibitem{}Moore, R. L., \& Stockman, H. S. 1981, \apj, 243, 60-75
\bibitem{}Morabito, D. D., Niell, A. E., Preston, R. A., Linfield, R. P.,
	Wehrle, A. E., \& Faulkner, J. 1986, \aj, 91, 1038-1050
\bibitem{}Morganti, R., Killeen, N. E. B., \& Tadhunter, C. N. 1993, \mnras, 263, 1023-1048
\bibitem{}Morton, D. C., Savage, A., \& Bolton, J. G. 1978, \mnras, 185, 735-740
\bibitem{}Neff, S. G., Hutchings, J. B., \& Gower, A. C. 1989, \aj, 97, 1291-1305
\bibitem{}Neugebauer, G., Miley, G. K., Soifer, B. T., \&  Clegg, P. E. 1986, \apj, 308, 815-828
\bibitem{}O'Dea, C. P., Baum, S. A., \& Stanghellini, C. 1991, \apj, 380, 66-77
\bibitem{}Padrielli, L., \etal 1986, \aap, 165, 53-75
\bibitem{}Pearson, T. J., \& Readhead, A. C. S. 1988, \apj, 328, 114-142
\bibitem{}Perley, R. A. 1982, \aj, 87, 859-880
\bibitem{}Peterson, B. A., Jauncey, D. L., Wright, A. E., \& Condon, J. J. 
	1976, \apj, 207, L5-L8
\bibitem{}Pica, A. J., Smith, A. G., Webb, J. R., Leacock, R. J., Clements, S.,
        \& Gombola, P. P. 1988, \aj, 96, 1215-1226
\bibitem{}Pian, E., \etal 1996, \apj, 459, 169-174
\bibitem{}Polatidis, A. G., \etal 1995, \apjs, 98, 1-32
\bibitem{}Preston, R. A., Morabito, D. D., Williams, J. G., Faulkner, J., 
	Jauncey, D.~L., \& Nicolson, G.~D. 1985, \aj, 90, 1599-1641
\bibitem{}Preston, R. A., \etal 1989, \aj, 98, 1-26
\bibitem{}Quiniento, Z. M., \& Cersosimo, J. C. 1993, \aaps, 97, 435-441
\bibitem{}Robertson, D. S., \etal 1993, \aj, 105, 353-358
\bibitem{}Rodgers, A. W., \& Peterson, B. A. 1977, \apj, 212, L9-L12
\bibitem{}Romero, G. E., Benaglia, P., \& Combi, J. A. 1995, \aap, 301, 33-40
\bibitem{}Romney,~J., \etal 1984, \aap, 135, 289-299
\bibitem{}Rusk, R. 1990, J. Roy. Astron. Soc. Can., 84, 199-215
\bibitem{}Schmidt, M. 1963, \nat, 197, 1040
\bibitem{}Shen, Z.-Q., \etal 1997, \aj, 114, 1999-2015 (Paper I)
\bibitem{}Shen, Z.-Q., Hong, X.-Y., \& Wan, T.-S. 1998, Acta Astrophysica Sinica, in press
\bibitem{}Shepherd,~M.~C., Pearson,~T.~J., \& Taylor,~G.~B. 1994, \baas, 26, 987
\bibitem{}Shimmins, A. J., Bolton, J. G., Peterson, B. A., \& Wall, J. V. 1971,
	\aplett, 8, 139-143
\bibitem{}Slee, O. B., Sadler, E. M., Reynolds, J. E., \& Ekers, R. D. 1994, \mnras, 269, 928-946
\bibitem{}Smith, H. J., \& Hoffleit, D. 1963, \nat, 198, 650-651
\bibitem{}Sparks, W. B., Miley, G. K., \& Macchetto, F.  1990, \apj, 361, L41-L44
\bibitem{}Steppe, H.,  H., Salter, C. J., Chini, R., Kreysa, E., Brunswig, W., \& 	
	P\'{e}rez, J. L. 1988, \aaps, 75, 317-351
\bibitem{}Steppe, H., Liechti, S., Mauersberger, R., K\"{o}mpe, C.,
        Brunswig, W., \& Ruiz-Moreno, M. 1992, \aaps, 96, 441-475
\bibitem{}Steppe, H., \etal 1993, \aaps, 102, 611-635 
\bibitem{}Stickel, M., Fried, J. W., \& K\"{u}hr, H. 1989, \aaps, 80, 103-114
\bibitem{}Stickel, M., Padovani, P., Urry, C. M., Fried, J. W., \& K\"{u}hr, H.
	1991, \apj, 374, 431-439
\bibitem{}Stickel, M., Fried, J. W., \& K\"{u}hr, H. 1993, \aaps, 98, 393-442
\bibitem{}Strittmatter, P. A., Serkowski, K., Carswell, R., Stein, W. A., Merrill, K. M.,
	\& Burbidge, E.~M. 1972, \apj, 175, L7-L13
\bibitem{}Swanenburg, B. N., \etal 1978, \nat, 275, 298
\bibitem{}Tanzi, E. G., \etal 1986, \apjl, 311, L13-L16
\bibitem{}Taylor, G. B., \etal 1994, \apjs, 95, 345-369
\bibitem{}Thakkar, D. D., \etal 1995, \apjs, 98, 33-40
\bibitem{}Thompson, D. J., \etal 1993a, \apj, 410, 87-89
\bibitem{}Thompson, D. J., \etal 1993b, \apjl, 415, L13-L16
\bibitem{}Thompson, D. J., \etal 1995, \apjs, 101, 259-286
\bibitem{}Tingay, S. J., \etal 1996, \apj, 464, 170-176
\bibitem{}Tornikoski, M., Valtaoja, E., Ter\"{a}sranta, H., Lainela, M.,
	Bramwell, D., \& Botti, L.~C.~L. 1993, \aj, 105, 1680-1689
\bibitem{}Treves, A., \etal 1993, \apj, 406, 447-450
\bibitem{}Ulvestad, J. S., Johnston, K. J., \& Weiler, K. W. 1983, \apj, 266, 18-27
\bibitem{}Unwin, S. C. \etal 1985, \apj, 289, 109-119
\bibitem{}Urry, C. M., \& Padovani, P. 1995, \pasp, 107, 803-845
\bibitem{}Vermeulen, R. C. \& Cohen, M. H. 1994, \apj, 430, 467-494
\bibitem{}von Montigny, C., \etal 1995, \apj, 440, 525-553
\bibitem{}Wall, J. V., Danziger, I. J., Pettini, M., Warwick, R. S., \& Wamsteker, W.
	1986, \mnras, 219, 23-29
\bibitem{}Wardle, J. F. C., Moore, R. L., \& Angel, J. R. P. 1984, \apj, 279, 93-111
\bibitem{}Webb, J. R., Smith, A. G., Leacock, R. J., Fitzgibbons, G. L.,
	Gombola, P. P., \& Shepherd, D. W. 1988, \aj, 95, 374-397
\bibitem{}Wehrle, A. E., Cohen, M. H., Unwin, S. C., Aller, H. D., Aller, M. F.,
	\& Nicolson, G. 1992, \apj, 391, 589-607
\bibitem{}Weiler, K. W., \& de Parter, I. 1983, \apjs, 52, 293-327
\bibitem{}Weiler, K. W., \& Johnston, K. J. 1980, \mnras, 190, 269-285
\bibitem{}White, G. L., \etal 1988, \apj, 327, 561-569
\bibitem{}Wilkes, B. J., Wright, A. E., Jauncey, D. L., \& Peterson, B. A. 1983, 
	Proc. Astron. Soc. Australia, 5, 2-9
\bibitem{}Wilkes, B. J. 1986, \mnras, 218, 331-361
\bibitem{}Wilkes, B. J., Elvis, M., Fiore, F., McDowell, J. C., Tananbaum, H., \&
	Lawrence, A. 1992, \apj, 393, L1-L4
\bibitem{}Wills, B. J., \& Browne, I. W. A. 1986, \apj, 302, 56-63
\bibitem{}Wills, B. J., Wills, D., Breger, M., Antonucci, R. R. J., \&
	Barvainis, R. 1992, \apj, 398, 454-475
\bibitem{}Wright, A., Peterson, B. A., Jauncey, D. L., \& Condon, J. J. 1979, \apj, 229, 73-77
\bibitem{}Wright, A., Ables, J. G., \& Allen, D. A. 1983, \mnras, 205, 793-807
\bibitem{}Xu, W., Readhead,~A.~C.~S., Pearson,~T.~J., Polatidis,~A.~G., \&
	Wilkinson,~P.~N. 1995, \apjs, 99, 297-348
\bibitem{}Zensus, J. A. 1987, in Superluminal Radio Sources,
	eds. J. A. Zensus and T. J. Pearson (Cambridge University Press, Cambridge), 26-31
\bibitem{}Zensus, J. A., B{\aa}{\aa}th, L. B., Cohen, M. H.,  \& Nicolson, G. D.  1988, \nat, 334, 410-412
\end{thebibliography}
\end{document}

\begin{deluxetable}{ccccclcl}
\pagestyle{empty}
\tablewidth{0pt}
\tablenum{3}
\tablecaption{Source Map Descriptions}
\tablehead{
&& \multicolumn{3}{c}{Restoring Beam} &&& \colhead{VLBI} \\
\colhead{Source} & \colhead{S$_{\small peak}$} & \colhead{Major} & \colhead{Minor} &
\colhead{P.A.} & \colhead{Contours} & \colhead{Scale} & \colhead{Morphology} \\
\colhead{Name} & \colhead{(Jy/Beam)}  & \colhead{(mas)}  & \colhead{(mas)} & \colhead{(deg)}
& \colhead{(Jy/Beam)} & \colhead{(pc/mas)} & \colhead{(mas scale)} \\
\colhead{(1)} & \colhead{(2)} & \colhead{(3)} & \colhead{(4)} & \colhead{(5)} & 
\colhead{(6)} & \colhead{(7)} & \colhead{(8)}}
\startdata
PKS~0118$-$272 & \phn 0.33 & 3.6 & 1.9 &  \phn $-$1  & 0.012$\times$($-$1,1,2,...,16)  & 3.7 & single core \\
PKS~0208$-$512 & \phn 2.38 & 2.2 & 1.1 &       $-$12 & 0.015$\times$($-$1,1,2,...,128) & 4.3 & core$-$jet \\
PKS~0332$-$403 & \phn 0.85 & 1.4 & 1.1 &       $-$10 & 0.006$\times$($-$1,1,2,...,64)  & 4.3 & core$-$jet \\
PKS~0403$-$132 & \phn 0.77 & 1.4 & 1.1 &  \phs    21 & 0.006$\times$($-$1,1,2,...,64)  & 3.7 & single core \\
PKS~0426$-$380 & \phn 0.67 & 1.7 & 1.2 &  \phs    16 & 0.006$\times$($-$1,1,2,...,64)  & 4.3 & single core \\
\\
PKS~0438$-$436 & \phn 1.08 & 1.7 & 1.0 &  \pq      4 & 0.015$\times$($-$1,1,2,...,64)  & 3.7 & core$-$jet \\
PKS~0454$-$234 & \phn 1.79 & 3.5 & 1.5 &  \phs    71 & 0.012$\times$($-$1,1,2,...,128) & 4.3 & core$-$jet \\
PKS~0521$-$365 & \phn 1.34 & 1.9 & 1.0 &  \phs    10 & 0.006$\times$($-$1,1,2,...,128) & 0.7 & core$-$jet \\
PKS~0537$-$441 & \phn 2.82 & 2.7 & 1.1 &  \phs    13 & 0.018$\times$($-$1,1,2,...,128) & 4.2 & core$-$jet \\
PKS~0823$-$223 & \phn 0.30 & 2.5 & 1.7 &  \pq      8 & 0.003$\times$($-$1,1,2,...,64)  & 4.2 & single core \\
\\
PKS~1034$-$293 & \phn 1.12 & 1.5 & 1.1 &  \phs    17 & 0.009$\times$($-$1,1,2,...,64)  & 2.8 & single core \\
PKS~1127$-$145 & \phn 1.35 & 2.8 & 1.4 &  \phs    69 & 0.009$\times$($-$1,1,2,...,128) & 4.3 & core$-$jet \\
PKS~1226+023   &     12.70 & 1.2 & 0.9 &  \phs    33 & 0.054$\times$($-$1,1,2,...,128) & 1.8 & core$-$jet \\ 
PKS~1244$-$255 & \phn 1.43 & 1.6 & 1.0 &  \pq      8 & 0.015$\times$($-$1,1,2,...,64)  & 3.9 & single core \\
PKS~1424$-$418 & \phn 0.96 & 2.3 & 1.1 &  \phn $-$2  & 0.012$\times$($-$1,1,2,...,64)  & 4.3 & core$-$jet \\
\\
PKS~1514$-$241 & \phn 1.10 & 2.0 & 1.0 &  \pq      5 & 0.024$\times$($-$1,1,2,...,32)  & 0.7 & single core \\
PKS~1936$-$155 & \phn 0.84 & 2.4 & 1.0 &  \pq      6 & 0.006$\times$($-$1,1,2,...,128) & 4.2 & single core \\
PKS~1954$-$388 & \phn 1.79 & 2.1 & 1.1 &  \phs    26 & 0.006$\times$($-$1,1,2,...,256) & 3.9 & single core \\
PKS~2005$-$489 & \phn 0.77 & 2.7 & 1.5 &  \phn $-$9  & 0.009$\times$($-$1,1,2,...,64)  & 0.9 & single core \\
PKS~2155$-$152 & \phn 1.48 & 2.1 & 1.4 &  \phs    62 & 0.015$\times$($-$1,1,2,...,64)  & 3.9 & core$-$jet \\
\\
PKS~2240$-$260 & \phn 0.35 & 1.5 & 1.1 &  \phs    10 & 0.006$\times$($-$1,1,2,...,32)  & 4.1 & core$-$jet \\
PKS~2243$-$123 & \phn 1.62 & 1.4 & 1.0 &  \pq      9 & 0.021$\times$($-$1,1,2,...,64)  & 3.9 & single core \\
PKS~2355$-$534 & \phn 1.31 & 2.8 & 1.2 &  \phs    20 & 0.009$\times$($-$1,1,2,...,128) & 4.3 & core$-$jet \\
\enddata
\tablecomments{\hfil\break
(1) Source name. \\
(2) Peak flux density.  \\
(3), (4), (5) Parameters of the restoring Gaussian beam:
	           Full Width at Half Maximum (FWHM) of the major and minor axes
                   and the position angle (P.A.) of the major axis. \\
(6) Contour levels of the map, starting from 3 times the rms value in the image
                   (`...' means a factor of 2 between adjacent contours). \\
(7) Angular size scale (with H$_0=100$ km~s$^{-1}$~Mpc$^{-1}$ and q$_0=0.5$). \\
(8) Apparent morphology of the source.}
\end{deluxetable}
\clearpage